\pdfoutput=1

\documentclass[11pt,twoside,a4paper,cmspaper,final,collab]{cms-tdr}
\def\svnVersion{280995}\def\cmsCernNo{2013-037}\def\cmsCernDate{\today}\def\cmsMessage{Published in Physics Letters B as \href{http://dx.doi.org/10.1016/j.physletb.2015.02.048}{\doi{10.1016/j.physletb.2015.02.048.}}}

\begin{document}\cmsNoteHeader{BPH-13-005}

\hyphenation{had-ron-i-za-tion}
\hyphenation{cal-or-i-me-ter}
\hyphenation{de-vices}

\RCS$Revision: 277002 $
\RCS$HeadURL: svn+ssh://svn.cern.ch/reps/tdr2/papers/BPH-13-005/trunk/BPH-13-005.tex $
\RCS$Id: BPH-13-005.tex 277002 2015-02-10 11:26:08Z argiro $
\cmsNoteHeader{BPH-13-005} 

\newcommand{\toprule}{\hline}
\newcommand{\midrule}{\hline}
\newcommand{\bottomrule}{\hline}
\newcommand{\usedlumi}{\ensuremath{20.7\fbinv\xspace}}
\newcommand{\usedlumiwitherror}{\ensuremath{20.7\pm 0.5\fbinv}\xspace}
\newcommand{\Chib}{\ensuremath{\chi_{\mathrm{b}}}\xspace}
\newcommand{\Chione}{\ensuremath{\chi_{\mathrm{b}1}}\xspace}
\newcommand{\Chitwo}{\ensuremath{\chi_{\mathrm{b}2}}\xspace}
\newcommand{\Chionetwo}{\ensuremath{\chi_\mathrm{b1,2}}\xspace}
\newcommand{\Chic}{\ensuremath{\chi_{\mathrm{c}}}\xspace}
\newcommand{\ChiCone}{\ensuremath{\chi_{\mathrm{c}1}}\xspace}
\newcommand{\ChiCtwo}{\ensuremath{\chi_{\mathrm{c}2}}\xspace}
\newcommand{\eratio}{\ensuremath{\varepsilon_{\Chione}/\varepsilon_{\Chitwo}}\xspace}
\newcommand{\theratio}{\ensuremath{N_{\Chitwo}/N_{\Chione}}\xspace}
\newcommand{\Y}{\ensuremath{\Upsilon}\xspace}
\newcommand{\YoneS}{\ensuremath{\Upsilon(1\text{S})}\xspace}
\newcommand{\YtwoS}{\ensuremath{\Upsilon(2\text{S})}\xspace}
\newcommand{\YthreeS}{\ensuremath{\Upsilon(3\text{S})}\xspace}
\newcommand{\psitwoS}{\ensuremath{\psi(2\text{S})}\xspace}
\newcommand{\roofit}{\texttt{RooFit}}

\title{Measurement of the production cross section ratio
       $\sigma(\Chitwo(1\mathrm{P}))/ \sigma(\Chione(1\mathrm{P}))$ in pp collisions at $\sqrt{s}= 8$\TeV}

\date{\today}

\abstract{
   A measurement  of the production cross section ratio $\sigma(\Chitwo(1\mathrm{P}))/\sigma( \Chione(1\mathrm{P}))$ is presented.
   The $\Chione(1\mathrm{P})$ and $\Chitwo(1\mathrm{P})$ bottomonium states, promptly
   produced in pp collisions at $\sqrt{s}= 8$\TeV, are detected by the CMS experiment at the CERN LHC through their radiative decays
   $\Chionetwo(1\mathrm{P}) \to \YoneS + \gamma$. The emitted photons
   are measured through their conversion to $\Pep\Pem$ pairs, whose reconstruction allows the two states to be resolved. The $\YoneS$ is measured through its decay to two muons.
   An event sample corresponding to an integrated luminosity of 20.7\fbinv  is used to measure the
   cross section ratio in a phase-space region defined by the photon pseudorapidity, $\abs{\eta^\gamma} < 1.0$;
   the \YoneS rapidity, $\abs{y^\Upsilon} < 1.5$; and the \YoneS transverse momentum, $7 < \pt^\Upsilon < 40$\GeV.
   The cross section ratio shows no significant dependence on the \YoneS transverse momentum, with a measured average value of $0.85 \pm 0.07\,\text{(stat+syst)}\pm 0.08\,\text{(BF)}$, where the first uncertainty is the combination of the experimental statistical and systematic uncertainties and the second is from the uncertainty in the ratio of the \Chib branching fractions.
}

\hypersetup{%
pdfauthor={CMS Collaboration},%
pdftitle={Measurement of the production cross section ratio sigma(chi[2b](1P))/sigma(chi[1b](1P)) in pp collisions at sqrt(s) = 8 TeV},%
pdfsubject={CMS},%
pdfkeywords={CMS, Quarkonium production, P-wave states, bottomonium}}

\maketitle 

\section{Introduction}
\label{sec:introduction}

Despite considerable efforts over the last decades, hadron formation, which is part of the nonperturbative sector of quantum chromodynamics (QCD),  remains poorly understood within the standard model of particle physics. Heavy-quarkonium production is an excellent probe of hadron formation.
In the past years, significant progress has been made in the theory sector~\cite{bib:Brambilla2011}, especially in the framework of nonrelativistic QCD (NRQCD)~\cite{bib:NRQCD}. This framework factorizes into  distinct processes the short-distance creation of a heavy quark-antiquark pair, in either a color-singlet or a color-octet configuration, and the long-distance formation of the quarkonium bound-state. The first process is presently calculated  to next-to-leading order in perturbative QCD~\cite{nlonrqcd}. The bound-state formation is described by transition probabilities, called long-distance matrix elements (LDMEs), which are assumed to be constant (independent of quarkonium transverse momentum  and rapidity) and universal (independent of the collision system and energy). In the Fock-state expansion of the heavy-quarkonium state, only a small number of color-singlet and color-octet terms contribute in the limit of small relative quark velocity v. The color-octet LDMEs are not easily calculable and the dominant ones are, therefore, treated as free parameters and adjusted to agree with the experimental data~\cite{bib:BK, bib:Chao,bib:GWWZ}.

The ratio of P-wave quarkonia production cross sections is a reliable test of predictions  because many theoretical, as well as experimental, uncertainties cancel out.
Prompt \Chic measurements in hadron collisions were not possible until the advent of  precise vertex detectors that allowed the separation of  promptly produced \Chic from those coming from the decay of B mesons~\cite{cdf:chi2chi1}. This ability is important, as NRQCD predictions are valid only for promptly produced \Chic. In the case of bottomonium, measurements are more difficult owing to the reduced production cross sections and the small separation in mass (19.4\MeV) between the $\Chione(1\mathrm{P})$ and the $\Chitwo(1\mathrm{P})$ (for readability the $1\mathrm{P}$ is dropped hereafter).
The production ratio of \ChiCtwo and \ChiCone is discussed in recent theoretical papers \cite{nlonrqcd,likhoded}, but the debate on the importance of color-octet contributions remains open. In the bottomonium sector, the NRQCD velocity expansion is more rigorously valid given the smaller relative quark velocity. Therefore, the measurement of the \Chitwo to \Chione production cross section ratio should give further insight into the mechanism that governs quarkonium production~\cite{likhoded}.
At the LHC, the charmonium $\ChiCtwo/\ChiCone$ production cross section ratio was measured by the LHCb~\cite{LHCb:2012ac}, CMS~\cite{cms:chi2chi1}, and ATLAS~\cite{ATLAS:2014ala} experiments, using data collected in $\Pp\Pp$ collisions at $\sqrt{s} = 7$\TeV. More recently, the LHCb experiment also reported a measurement of the $\Chione$ and $\Chitwo$ production cross section ratio using combined $\sqrt{s} = 7$ and  8\TeV data~\cite{LHCB:2014chib}.

This Letter presents a measurement of the $\Chitwo/\Chione$ production cross section ratio.
The $\Chione$ and $\Chitwo$
states are reconstructed by detecting their radiative decays
$\chi_{\mathrm{b}1,2} \to \Upsilon(1\mathrm{S}) + \gamma$, which is the dominant decay mode, with the \YoneS decaying into two muons.
An accurate measurement of the photon energy (typically in the range 0.5--2\GeV) is obtained from
the reconstruction of the momentum of the electron-positron pair originating from the photon conversion in the
beam pipe or in the inner layers of the CMS silicon tracker. The resulting mass resolution  of the $\chi_\mathrm{b}$ candidates, around 5\MeV, is sufficient to resolve the two
$\chi_\mathrm{b1,2}$ peaks at the expense of a limited yield, given the small reconstruction efficiency for such low-energy photons.
The cross section ratio is obtained as
\begin{linenomath}
\begin{equation}
\label{eq:rp}
{\mathcal{R}} \equiv
\frac{\sigma(\Pp\Pp \to \Chitwo + X)}
       {\sigma(\Pp\Pp \to \Chione + X)} =
\frac{N_{\Chitwo}}{N_{\Chione}} \cdot
\frac{\varepsilon_{\Chione}}{\varepsilon_{\Chitwo}} \cdot
\frac{\mathcal{B}(\Chione\to \YoneS + \gamma)}
        {\mathcal{B}(\Chitwo\to \YoneS + \gamma)},
\end{equation}
\end{linenomath}
where $N_{\chi_\mathrm{b1,2}}$ are the yields of $\chi_{\mathrm{b}1,2}$ signal candidates,
simultaneously obtained from an unbinned maximum likelihood fit of the $\mu\mu\gamma$
invariant-mass spectrum, \eratio is the ratio of the acceptance and efficiency corrections for the two processes obtained from a full detector
simulation, and $\mathcal{B}(\chi_\mathrm{b1,2}\to \YoneS + \gamma)$ are the branching
fractions of the corresponding radiative decays~\cite{PDG}.
The results are presented in four bins of \YoneS transverse momentum, $\pt^\Upsilon$, in the range 7--40\GeV. This choice was driven by the trigger requirements at low \pt, and the amount of available data at high \pt.

\section{CMS detector and event samples}

The central feature of the CMS apparatus is a superconducting solenoid of 6\unit{m} internal diameter,
providing a magnetic field of 3.8\unit{T}. Within the superconducting solenoid volume are a silicon pixel
and strip tracker, a lead tungstate crystal electromagnetic calorimeter, and a brass/scintillator
hadron calorimeter. Muons are measured in the pseudorapidity range $\abs{\eta^\mu}< 2.4$ in
gas-ionization detectors embedded in the steel flux-return yoke outside the solenoid.
The first level of the CMS trigger system, composed of custom hardware processors, uses information
from the calorimeters and muon detectors to select the most interesting events. The
high-level trigger processor farm further decreases the event rate  before data storage.
A more detailed description of the CMS detector, together with a definition of the coordinate system used and the relevant kinematic variables, can be found in Ref.~\cite{Chatrchyan:2008zzk}.

The analysis is based on the $\sqrt{s} = 8$\TeV $\Pp\Pp$ data sample
collected by CMS in 2012 at the CERN LHC, corresponding to an integrated luminosity of \usedlumi.
The events have been selected at the trigger level by requiring opposite-sign muon pairs of invariant mass
in the range 8.5--11.5\GeV, dimuon \pt larger than 6.9\GeV, a distance of closest approach of each muon track to the beam axis of
less than 5\unit{mm}, and a $\chi^2$ probability from the kinematic fit of the muons to a common vertex larger than 0.5\%.

To parameterize the reconstructed  $\chi_\mathrm{b1,2}$ mass distributions and evaluate the reconstruction efficiency, a detailed  Monte
Carlo (MC) simulation based on \GEANTfour \cite{Geant4_2} was performed. About 40~million events for each \Chib
state were propagated through an accurate description of the CMS detector, including realistic trigger
emulations and reconstruction algorithms identical to those used to process the collected data.
Since the photon conversion probability multiplied by the reconstruction efficiency is less than 1\% for photons of energy below 1\GeV, a large number of events is needed. To reduce CPU usage, the $\Chib$ mesons were generated alone, without any ``underlying event''.
This simplification should have no influence on the results of the analysis because any possible effect of accompanying particles on the track reconstruction efficiencies is identical for the two $\Chib$ states, cancelling in the ratio.
Both $\Chib$ samples were produced using the \PYTHIA event generator~\cite{Sjostrand:2006za}, with the $\Chib$ \pt distributions parameterized on the basis of the
\YtwoS \pt spectrum measured by CMS~\cite{cms:Ypapers}.
The \Chib mesons were generated in the rapidity range $\abs{y} < 2.0$ and forced to decay into
$\YoneS + \cPgg$, with the \YoneS mesons decaying to dimuons.
Only simulated events where a photon conversion occurred were further processed and reconstructed.

\section{Event reconstruction and selection}
\label{sec:reco_selection}

The CMS muon reconstruction procedure~\cite{cms:muons} identifies muons by requiring that
tracks reconstructed in the silicon tracker be matched with at least one muon segment in any muon detector.
To ensure an accurate \pt measurement and to suppress the contribution from decays-in-flight of pions and kaons, the
number of silicon tracker layers with at least one hit must be larger than five, with two of them in the silicon
pixel layers, and the track-fit $\chi^2$ per degree of freedom  must be smaller than~1.8.
Loose selections are applied to the transverse and longitudinal muon impact parameters, $d_{xy} < 3$\cm and
$\abs{d_z} < 30$\cm, respectively, to further suppress decays in flight and cosmic ray muons.
The selected muons must have a transverse momentum  $\pt^\mu$  and pseudorapidity $\eta^\mu$ in the fiducial kinematic region $\pt^\mu > 3.5$\GeV and $\abs{\eta^\mu} < 1.9$.
Each event containing a pair of opposite-sign muons is kept in the analysis sample if the dimuon invariant mass $m_{\mu\mu}$ is between 8.5
and 11\GeV, its absolute rapidity is less than 1.5, and the $\chi^2$ probability of the dimuon kinematic fit (with the
two muon tracks constrained to a common vertex) is larger than 1\%.
To select \YoneS candidates, $m_{\mu\mu}$  is required to be within $3 \sigma$ of the \YoneS mass, where the dimuon mass resolution $\sigma$ is parameterized as a function of the \YoneS rapidity, and is obtained by fitting the dimuon mass distribution in narrow dimuon rapidity bins.
This parameterization accounts for the significant rapidity dependence of the dimuon mass resolution, from around 65\MeV at $y=0$ to around 120\MeV at $\abs{y} = 1.5$.

The low-energy photons produced in the \Chib radiative decays that convert into electrons and
positrons often produce tracks that are rather asymmetric, with one of the two tracks carrying a small fraction of the photon's energy. Given that these tracks rarely reach the calorimeter, they are
reconstructed exclusively using information from the silicon tracker.
An algorithm optimized for the reconstruction of low-\pt displaced tracks has been used,
relying on an iterative tracking procedure~\cite{trackernew}.

The track pairs used to reconstruct the converted photons are required to fulfill the following selection criteria:
the two tracks must be of opposite charge;
one must have at least four hits in the silicon tracker layers and the other at least three hits;
the innermost hits of the two tracks must be less than 5\cm apart along the beam direction;
both tracks must have a reduced track-fit $\chi^2$ smaller than 10;
the two tracks should be almost parallel to each other, having angular separations
$\Delta(\cot\theta) < 0.1$ and $\Delta\phi < 0.2$ rad,
where $\theta$ and $\phi$ are the polar and azimuthal angles, respectively, defined at the common vertex; the primary \Pp \Pp~vertex associated with the photon conversion is required to lie outside both track helices;
defining  $d_m$ as the distance between the centers of the two circles formed by the tracks in the transverse plane  minus the sum of their radii, the condition  $-0.25 < d_m < 1$\cm must be satisfied;
finally, the conversion vertex must be at least 1.5\cm away from the beam axis in the transverse plane, in order to suppress the
background from $\pi^0$ Dalitz decays.

The analysis is  performed on events having a photon with pseudorapidity $\abs{\eta^\gamma} < 1.0$,
where the photon energy is measured with the best resolution, an important factor for a clean separation between
the two \Chib peaks.
When the \YoneS, loosely selected as described in above, and converted photons are paired to form \Chib candidates,
the distance along the beam axis between the dimuon vertex and the extrapolated photon trajectory is required
to be less than 1\mm.
The invariant mass of the \Chib candidate, $m_{\mu\mu\gamma}$, is calculated through a kinematic fit,
which constrains
the dimuon invariant mass to the \YoneS mass and the electron-positron invariant mass to zero.
In addition, the electron and positron are constrained to a common vertex, as are the two muons and the photon.
The \Chib candidates are retained if the $\chi^2$ probability of the kinematic fit is larger than 2\%.
This approach significantly reduces the effect of the muon momentum resolution on the \Chib mass resolution.

\section{Analysis procedure}
\label{sec:signalextraction}

The shape of the reconstructed invariant-mass distributions of the \Chib candidates is evaluated through MC
simulation, performed under the assumption that
the intrinsic width of the \Chib states, predicted to be smaller than 1\MeV~\cite{chiwidths}, is negligible compared to the mass resolution, which is of the order of 5\MeV.
Since the low-mass tail of the \Chitwo shape falls under the \Chione peak, it is important to use a reliable
parameterization of the resolution function when evaluating the ratio of the \Chitwo and \Chione yields, \theratio.
The \Chib mass resolution is dominated by the energy resolution of the converted photon and
has a clear low-mass tail, typical of processes involving radiative losses (the electrons and positrons
lose energy when traversing the tracker material).
The simulated signal shape also reveals the presence of a small high-mass tail due to multiple scattering; the signal response is parameterized by a double-sided Crystal Ball (CB) function~\cite{xball} consisting of a Gaussian core with two power-law tails, with independent exponents and transition points.

\begin{figure*}[bth!]
    \includegraphics[width=0.45\linewidth]{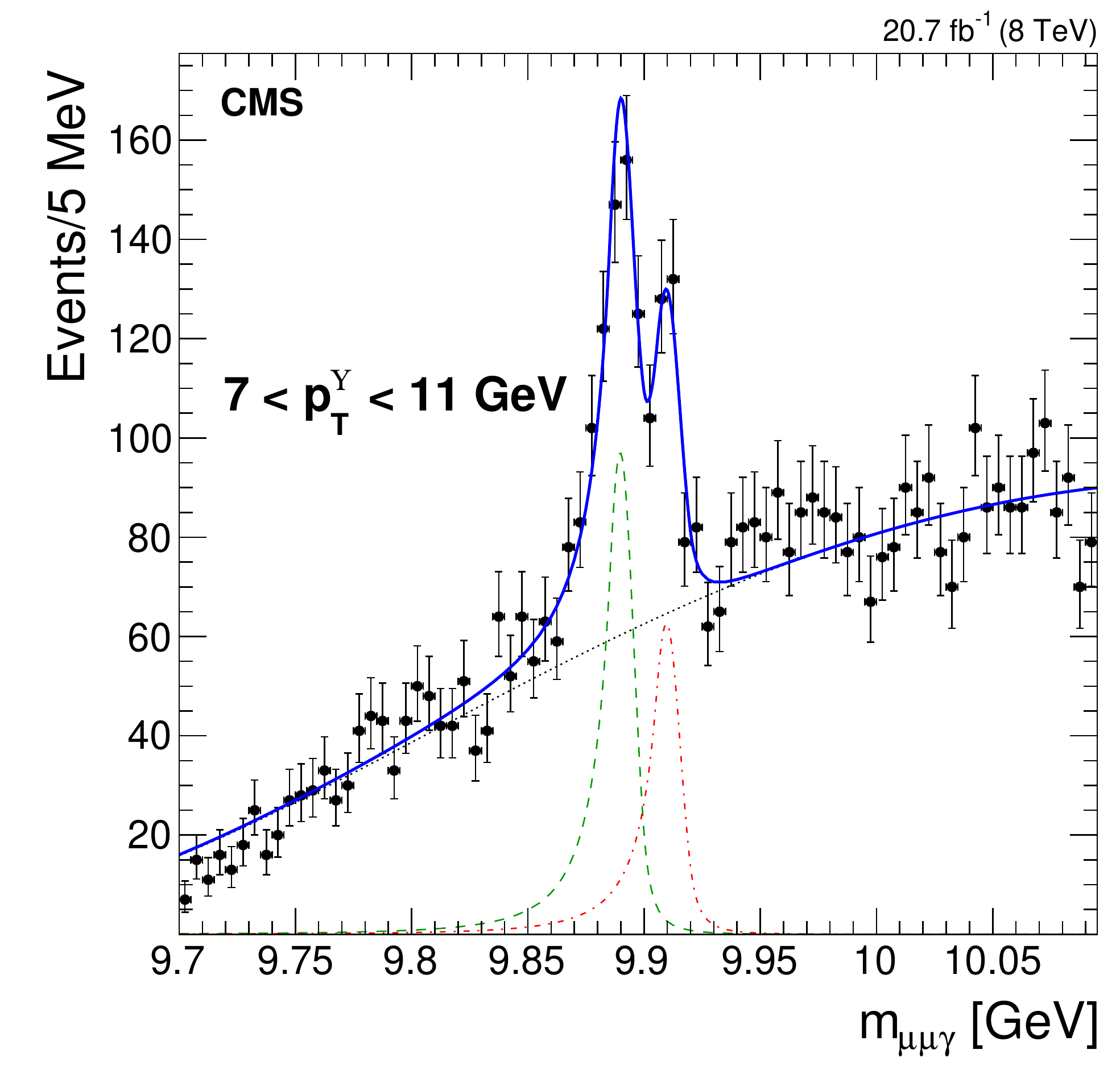}
    \includegraphics[width=0.45\linewidth]{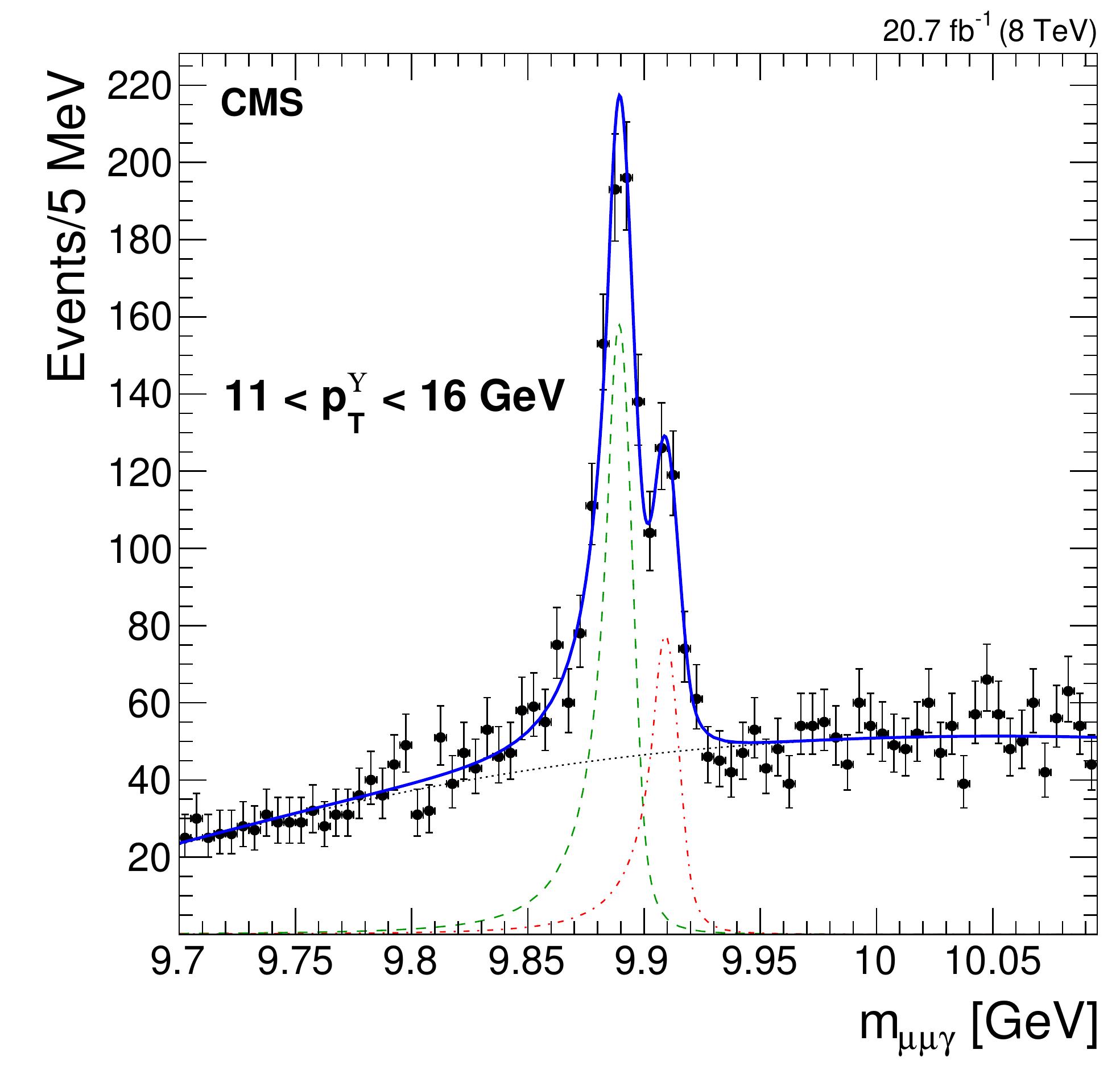}\\
    \includegraphics[width=0.45\linewidth]{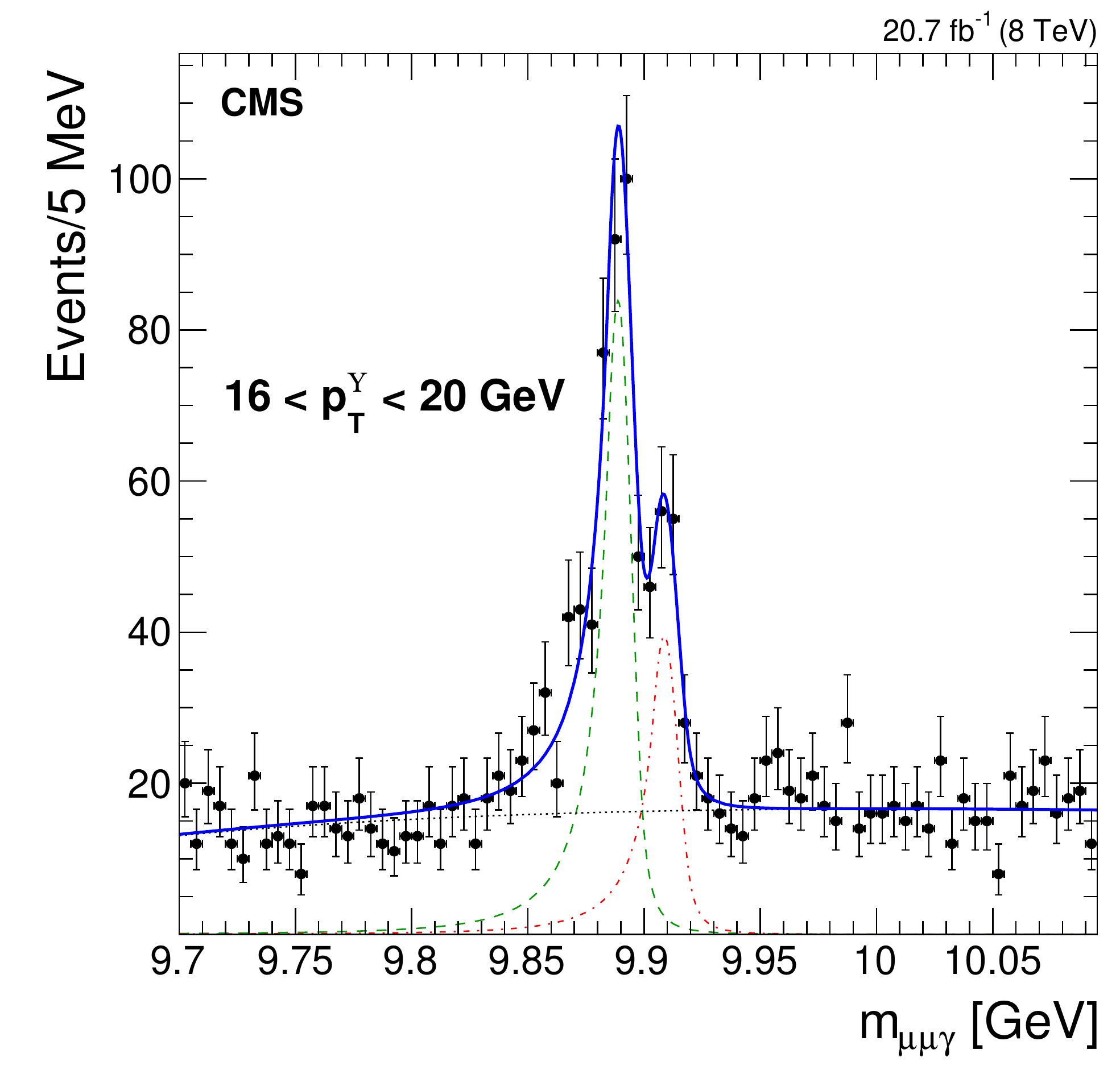}
    \includegraphics[width=0.45\linewidth]{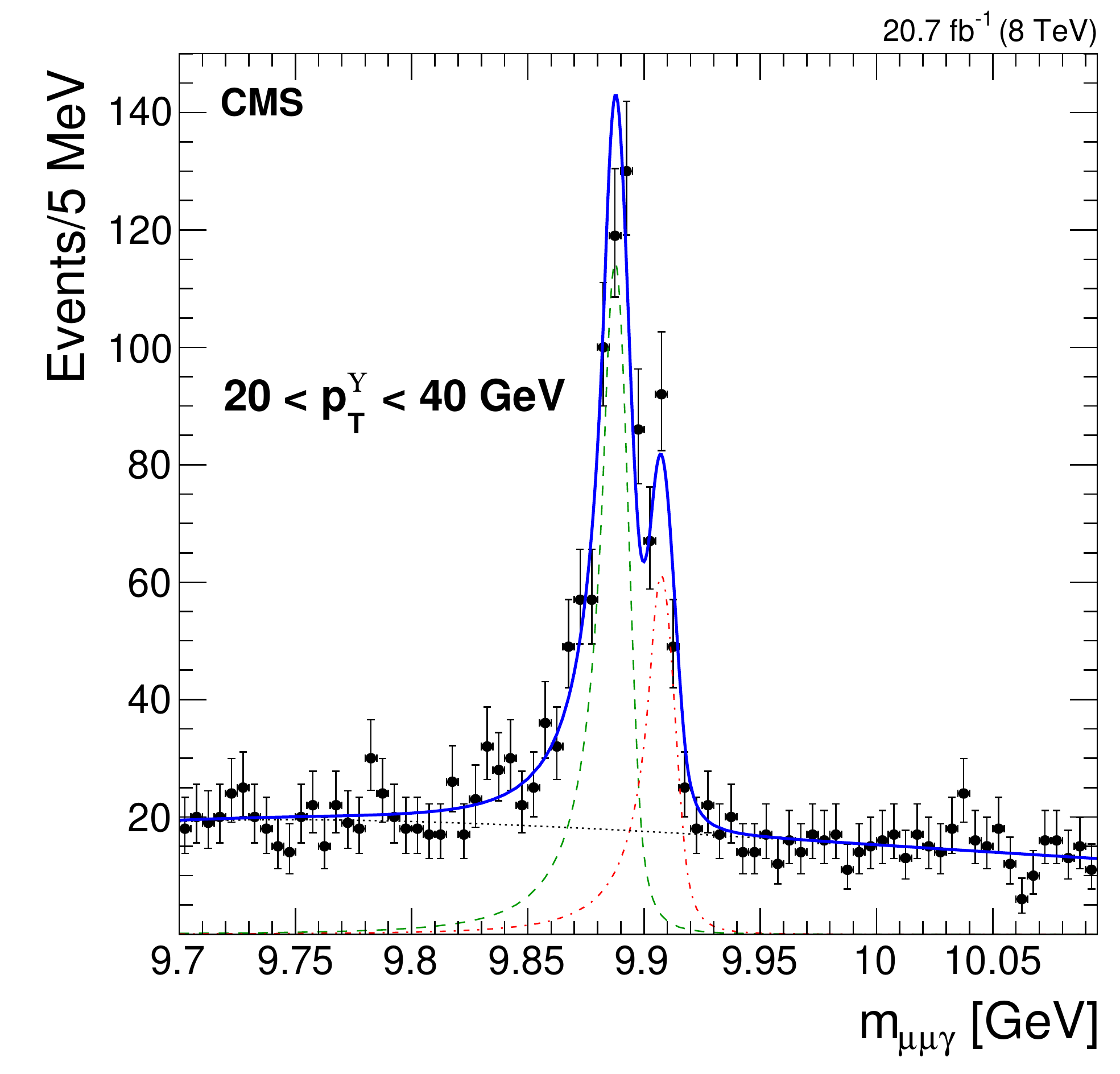}
    \caption{Invariant-mass distributions of the $\mu\mu\gamma$ candidates for each of the four \YoneS \pt bins considered in
    the analysis. The fitted \Chione and \Chitwo signals
    are parameterized with double-sided CB functions determined using simulated events. The combinatorial background is described by the product of an exponential and a power-law function. The solid line gives the result of the overall fit, with the dashed and dashed-dotted lines showing the \Chione and \Chitwo contributions, respectively. The dotted line represents the background contribution.}
    \label{fig:masspectra}
\end{figure*}

The ratio \theratio is measured with an unbinned maximum likelihood fit to the $\mu\mu\gamma$ invariant-mass distribution, in four bins of \YoneS \pt.
The \Chione and \Chitwo probability density functions are modelled by double-sided CB functions with shape parameters fitted to the simulated distributions in each  \pt bin of the \YoneS. The total \Chione yield, the \theratio ratio, and the total number of background events $N_\text{bkg}$ are free parameters when fitting the data.
The underlying continuum background, composed predominantly of events where the \YoneS and photon are unrelated,
is modelled by a probability distribution function proportional to $(m-m_0)^\lambda \cdot \exp({-\nu(m-m_0)})$,
where $m$ is the $\mu\mu\gamma$ invariant mass obtained from the four-track kinematic fit, $m_0 = 9.5$\GeV,
and $\lambda$ and $\nu$ are free parameters. The fit is performed in the $\mu\mu\gamma$ mass region 9.7--10.1\GeV .
Figure~\ref{fig:masspectra} shows the fitted invariant-mass distributions from data  for each of the four \YoneS \pt bins considered, while
Table~\ref{table:fitresults} gives the corresponding fit results.

\begin{table*}[tbh]
\topcaption{Parameters extracted from the fits to the measured $\mu\mu\gamma$ invariant mass distributions,
in the four \YoneS \pt bins considered. The quoted uncertainties are statistical only. The parameters are defined in the text.}
\label{table:fitresults}
\centering
 \begin{tabular}{c*{4}{c}}
\toprule
\rule{0pt}{2ex} $\pt^\Upsilon$ [\GeVns{}]  & 7--11  & 11--16  & 16--20  & 20--40 \\
\midrule
$N_{\Chitwo}/N_{\Chione}$ & $0.65 \pm 0.12$ & $0.51 \pm 0.06$ & $0.49 \pm 0.08$ & $0.54 \pm 0.07$\\
$N_{\Chione}$ & $392 \pm 37$ & $655 \pm 39$ & $343 \pm 27$ & $474 \pm 31$\\
$N_\text{bkg}$ & $5772 \pm 86$ & $3909 \pm 72$ & $1401 \pm 43$ & $1431 \pm 44$\\
$\lambda$ & $3.50 \pm 0.23$ & $2.08 \pm 0.45$ & $0.78 \pm 0.72$ & $0.79 \pm 0.71$\\
$\nu$ [\GeVns{}$^{-1}$] & $6.01 \pm 0.66$ & $4.09 \pm 1.17$ & $1.59 \pm 1.80$ & $3.19 \pm 1.85$\\

\bottomrule
 \end{tabular}
\end{table*}

The measurement of the cross section ratio $\mathcal{R}$, defined in Eq.~(\ref{eq:rp}), depends on the ratio of the
\Chione and \Chitwo measurement acceptance and efficiencies, $\varepsilon_{\Chione}/\varepsilon_{\Chitwo} =
( N^\text{rec}_{\Chione}/N^\text{gen}_{\Chione} ) \big/
( N^\text{rec}_{\Chitwo}/N^\text{gen}_{\Chitwo} )$,
where, for each $\pt^\Y$ bin, $N^\text{rec}$ is the number of simulated candidates that are reconstructed and pass the
event selection criteria, while $N^\text{gen}$ is the corresponding number of generated candidates, in the
kinematic window $\abs{y^\Y} < 1.5$, $\pt^\mu > 3.5$\GeV, $\abs{\eta^\mu} < 1.9$, and $\abs{\eta^\gamma} < 1.0$.
The reconstructed kinematic distributions of the simulated decay products are found to be in agreement with the measured ones.
The values of the acceptance and efficiency ratios, \eratio, are shown in Table~\ref{table:eratio}. The ratio of the acceptances times efficiencies differs from unity owing to the increased detection efficiency of the higher-energy  photon from the \Chitwo decay.

\begin{table*}[tbh]
\topcaption{Acceptance and efficiency ratios, \eratio, computed from MC simulation. Uncertainties are statistical only.}
\label{table:eratio}
\centering
\begin{tabular}{c*{4}{c}}
\toprule
\rule{0pt}{2ex}$\pt^\Upsilon$ [\GeVns{}]  & 7--11  & 11--16  & 16--20  & 20--40 \\
\midrule
$\varepsilon_{\Chione}/\varepsilon_{\Chitwo}$ & $0.85 \pm 0.05$  & $0.91 \pm 0.05$  & $0.91 \pm 0.07$  & $0.92 \pm 0.04$ \\
\bottomrule
 \end{tabular}
\end{table*}

\section{Systematic uncertainties
\label{sec:systematics}}

The \Chione and \Chitwo signal shapes are derived from MC simulation.
To evaluate the uncertainty in the \theratio ratio stemming from the imperfect parameterization of the MC signal shape,
caused by the finite number of simulated events,
a large number of pseudo-experiments are generated, randomly drawing sets of shape parameters using the covariance matrices of the fits to the simulated distributions. They are then used to fit the measured mass distributions, and the resulting \theratio distribution is fitted with a Gaussian
function, the  standard deviation of which is taken as the systematic uncertainty corresponding to the ``signal parameters''.

To account for possible discrepancies between the simulated and measured events regarding, in particular, the energy scale
calibration and the measurement resolution, alternative data-fitting schemes are used, leaving some of the signal shape parameters
free in the fit to the measured mass distributions.
A Chebyshev polynomial function is also used as an alternative model for the shape of the mass distribution of the background.
The maximum variation in the \theratio ratio with these different fitting strategies is taken as the ``signal and background modelling"
systematic uncertainty.
The fitting procedure is found to be unbiased, as judged using pseudo-experiments where a certain \theratio value is injected; the fitted
results deviate on average from the input values by less than 10\% of the statistical uncertainty.
It has also been verified that the \theratio ratio is insensitive to the addition of a signal term describing the $\chi_\mathrm{b0}$
state.
The possible influence of multiple primary vertices in the event (``pileup") on the \theratio measurement was
investigated by repeating  the analysis in subsamples of events with different numbers of reconstructed primary vertices. No statistically significant effect is observed.

Another source of systematic uncertainty in the evaluation of $\mathcal{R}$ is the
statistical uncertainty in \eratio, reflecting the finite size of the simulated event samples used to evaluate the measurement acceptances and efficiencies.
The influence of the generated \Chione and \Chitwo \pt spectra on \eratio is evaluated by using alternative functions instead of the \YtwoS spectrum.
A reweighting procedure is used to obtain the values of \eratio corresponding to scenarios where the \Chione and
\Chitwo are both produced according to the \YoneS or \YthreeS \pt spectra, as well as to mixed scenarios where the two states
have different spectra: \YoneS for the \Chione and \YtwoS for the \Chitwo, or \YtwoS for the \Chione and \YthreeS for the \Chitwo.
The maximum variation in the \eratio values obtained with all hypotheses is taken as the \Chib \pt spectra uncertainty.
Possible dependencies of the \eratio determination on the description of the tracking detector material in the MC simulation have been found to be negligible.

The acceptances and efficiencies are evaluated under the assumption that the \Chione and \Chitwo are both produced unpolarized.
Polarization affects the angular and \pt distributions of the radiated photon;
since the photon reconstruction efficiency significantly depends on the photon \pt, especially at low \pt, the ratio of
acceptances and efficiencies depends on the polarization scenario.
In order to investigate this effect, the unpolarized MC distributions are reweighted to reproduce  the
theoretical angular distributions of \Chib decay products expected for different \Chib polarizations~\cite{facciolimc}.
The acceptance and efficiency ratio is recalculated assuming that the \Chione is produced unpolarized or with helicity $m_{\Chione} = 0, \pm 1$,
in combination with the assumption that  the $\Chitwo$ is produced unpolarized or with helicity $m_{\Chitwo} = 0, \pm1, \pm 2$, both
in the helicity and Collins--Soper~\cite{cs} frames.
The maximal variations of \eratio in these $4\times 6$ scenarios with respect to the ``both unpolarized" case
have a negligible influence (at the percent level) on the cross section ratio, well below the other uncertainties.
Table~\ref{table:systematics} summarizes the systematic uncertainties considered in the analysis.

\begin{table*}[htb]
\topcaption{Relative systematic uncertainties in
$\frac{N_{\Chitwo}}{N_{\Chione}} \cdot \frac{\varepsilon_{\Chione}}{\varepsilon_{\Chitwo}}$, in percent,
for the four $\pt^\Y$ bins considered.}
\label{table:systematics}
\centering
\begin{tabular}{lcccc}
\toprule
\rule{0pt}{2ex}$\pt^\Upsilon$ [\GeVns{}]  & 7--11  & 11--16  & 16--20  & 20--40 \\ \midrule
Signal parameters                & 3.2  & 3.5 & 3.3 & 3.5\\
Signal and background modelling  & 6.8  & 5.3 & 5.2 & 6.1\\
\eratio statistical uncertainty  & 6.4  & 5.8 & 7.2 & 4.4\\
Choice of \Chib \pt spectra      & 3.0  & 3.0 & 2.8 & 4.8\\  \midrule
Total                            & 10.3 & 9.1 & 9.9 & 9.6 \\
\bottomrule
 \end{tabular}
\end{table*}

\section{Results and discussion}
\label{sec:results}

The ratio $\mathcal{R}$ of the
\Chitwo and \Chione production cross sections in the $\Upsilon(1\mathrm{S}) + \gamma$ decay channel for each $\pt^\Upsilon$ bin is obtained by correcting the
$N_{\Chitwo}/N_{\Chione}$ yield ratio (Table~\ref{table:fitresults})
with the corresponding  acceptance and efficiency ratio $\varepsilon_{\Chione}/\varepsilon_{\Chitwo}$ (Table~\ref{table:eratio}).

\begin{table*}[tbh]
\topcaption{The cross section ratio $\sigma({\Chitwo}) \,/\, \sigma({\Chione})$ measured in four $\pt^\Y$ bins
before and after multiplying by  the $\Chib \to \Upsilon(1S)+ \gamma$ branching fractions~\cite{PDG}. The second column gives the average \pt value for each bin. The first uncertainty is statistical, the second is systematic, and the third reflects the uncertainty in the branching fraction.}
\label{table:results}
\centering
\begin{tabular}{cccc}
\toprule
\rule{0pt}{2ex}$\pt^\Upsilon$ [\GeVns{}]  &
$\langle \pt^\Upsilon \rangle$ [\GeVns{}]
& $\sigma(\Chitwo)\,/\,\sigma(\Chione) \times  \mathcal{B}(\Chitwo) \,/\, \mathcal{B}(\Chione) $
&  $\sigma(\Chitwo)\,/\,\sigma(\Chione)$  \\
\midrule
 7--11 & 8.7& 0.56 $\pm$ 0.10  $\pm$ 0.06  & 0.99 $\pm$ 0.18  $\pm$ 0.10  $\pm$ 0.09  \\
11--16 & 12.9& 0.47 $\pm$ 0.06  $\pm$ 0.04  & 0.83 $\pm$ 0.10  $\pm$ 0.08  $\pm$ 0.07  \\
16--20 & 17.5& 0.45 $\pm$ 0.07  $\pm$ 0.04  & 0.80 $\pm$ 0.13  $\pm$ 0.08  $\pm$ 0.07  \\
20--40 & 26.2& 0.50 $\pm$ 0.06  $\pm$ 0.05  & 0.89 $\pm$ 0.11  $\pm$ 0.09  $\pm$ 0.08  \\
\bottomrule
\end{tabular}
\end{table*}

\begin{figure*}[h!tbp]
 \centering
 \includegraphics[width=0.49\linewidth]{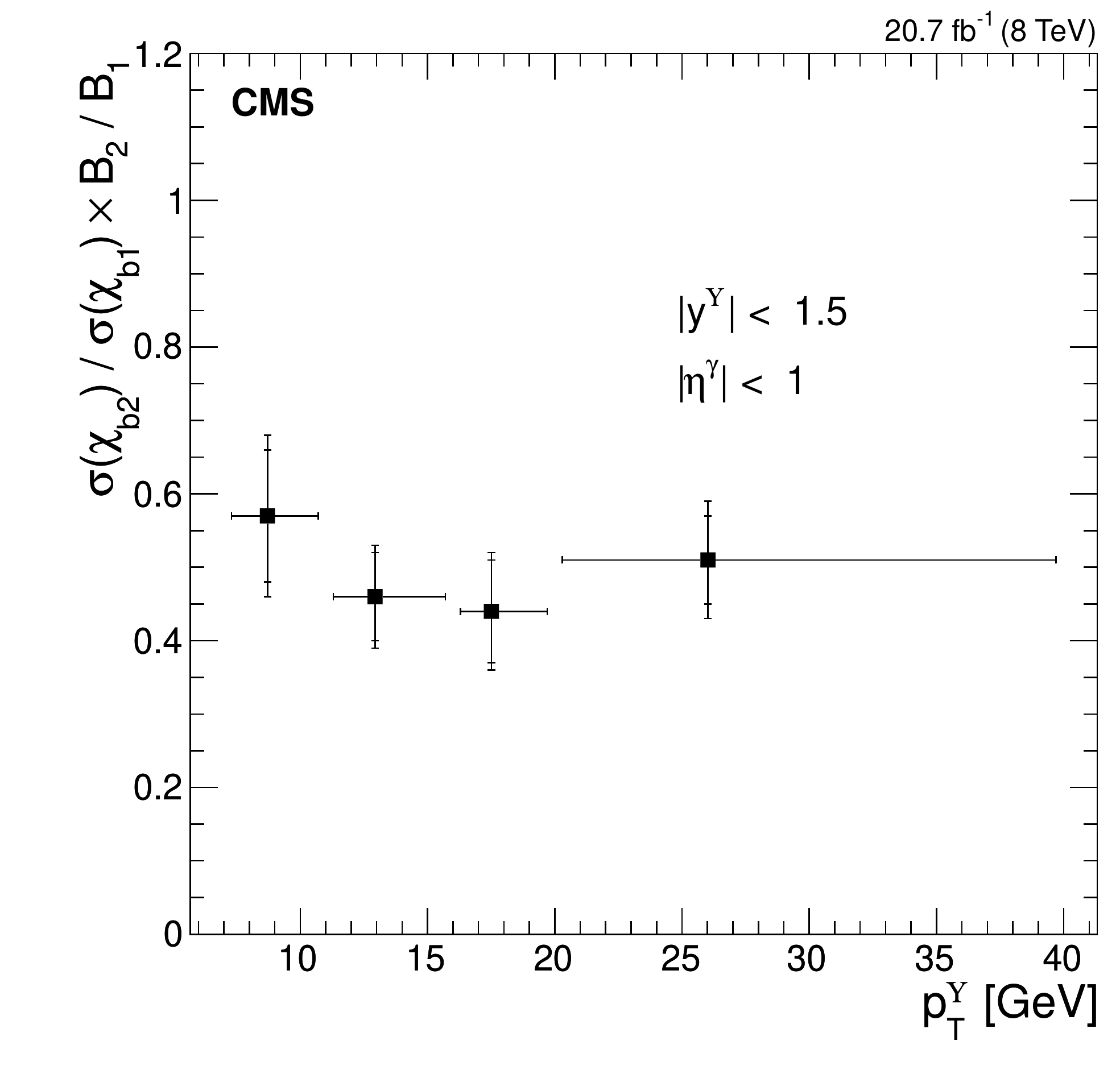}
 \includegraphics[width=0.49\linewidth]{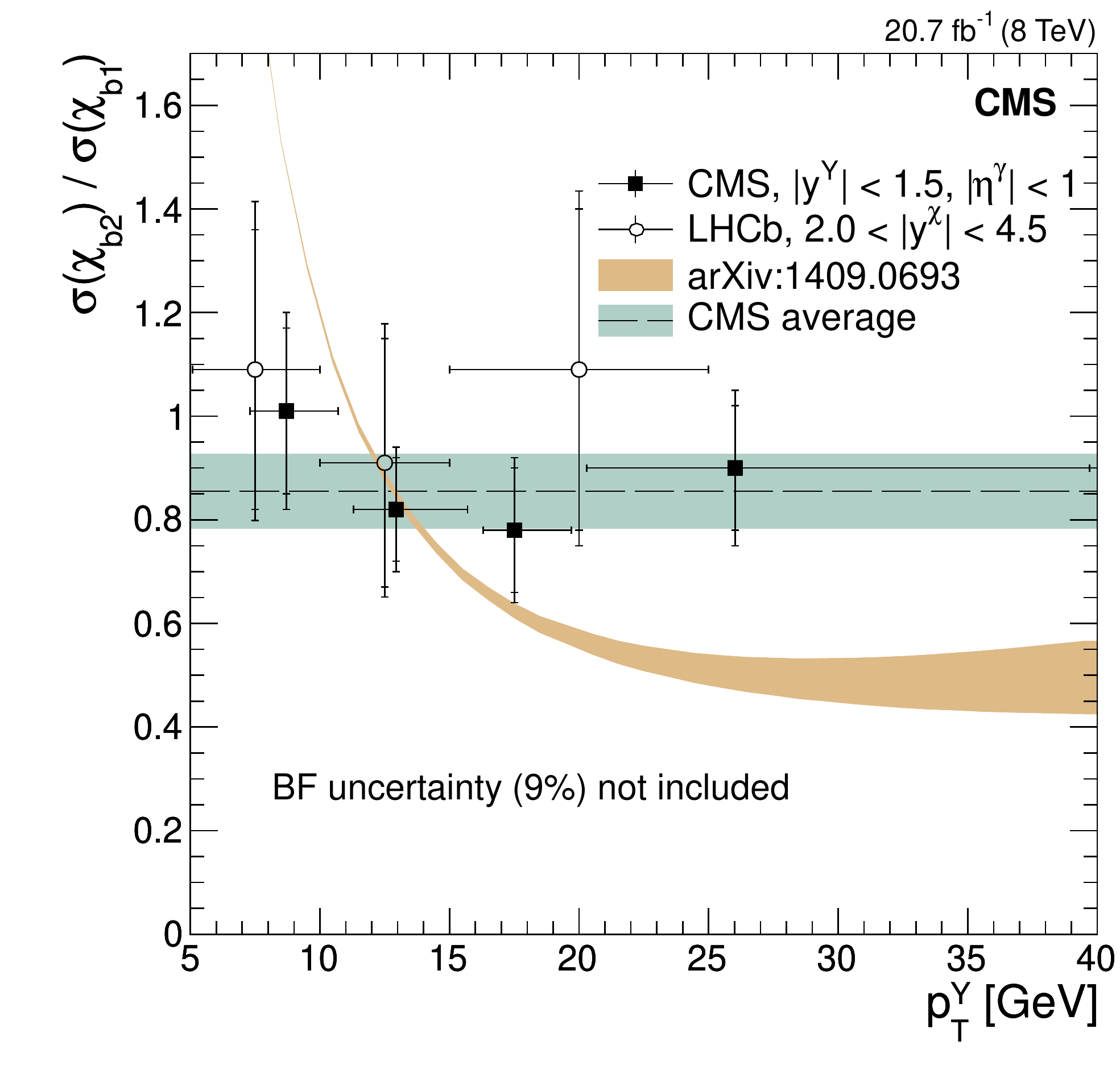}
\caption{The ratio of the $\Chitwo$ and $\Chione$ production cross sections, as a function of $\pt^\Y$, before (left)
and after (right) multiplying by the ratio of the $\YoneS + \cPgg$ branching fractions, as measured by CMS and LHCb~\cite{LHCB:2014chib} (right only).
The vertical bars represent the statistical (inner bars) and total (outer bars) experimental uncertainty, respectively. The horizontal bars show the width of each bin. The dashed line in the right plot is a fit of a constant to the CMS measurements, and the horizontal band is the total uncertainty in the fit result. The 9\% uncertainty in the ratio of branching fractions, which applies to all bins of $\pt^\Y$, is not included.
The curved band represents the result of a theoretical calculation~\cite{stanislav}.}
\label{fig:results}
\end{figure*}

Figure~\ref{fig:results} shows the measured $\sigma({\Chitwo})/\sigma({\Chione})$ cross section ratio,
as a function of $\pt^\Y$, before (left)  and after (right) multiplying by the decay branching fractions, taken from Ref.~\cite{PDG}. The relative uncertainty in the ratio of the branching fractions is 9\%. The numerical values are given in Table~\ref{table:results}.

The open circles in Fig.~\ref{fig:results}, right panel, show the LHCb measurement \cite{LHCB:2014chib}, which can be compared to the CMS result since no significant dependence of the ratio on the rapidity is expected.
The shaded area in the right panel of Fig.~\ref{fig:results} shows a theoretical calculation \cite{stanislav} performed in the framework of NRQCD. Since data on \Chib production were not available until recently, in this calculation the LDMEs are extracted from experimental data on the $\sigma({\ChiCtwo})/\sigma({\ChiCone})$ cross section ratio~\cite{cms:chi2chi1, LHCb:2012ac, cdf:chi2chi1} and extrapolated, using NRQCD scaling rules, to the case of P-wave bottomonium.  The dashed line in Fig.~\ref{fig:results} (right) gives the result of a fit of the CMS measurements to a constant, corresponding to $0.85 \pm 0.07$, where the uncertainty includes both statistical and systematic uncertainties, but not the uncertainty in the ratio of the \Chib branching fractions. A constant behavior is expected in the case of color-octet dominance. The measurements do not  indicate the large increase in the ratio at low $\pt^\Upsilon$ and  differ by more than two standard deviations from the asymptotic value at high \pt predicted by the theory. More precise measurements may be needed in order to thoroughly test the validity of NRQCD in the P-wave bottomonium sector.

\section{Summary}
\label{sec:summary}

The production cross section ratio $\sigma(\Chitwo(1\mathrm{P}))/\sigma( \Chione(1\mathrm{P}))$ has been measured in \Pp \Pp~collisions by detecting the radiative decays to an \YoneS and a photon, with the \YoneS  decaying to two muons. Events are selected where the \YoneS and photon are emitted in the phase-space region defined by $\abs{y^\Upsilon} < 1.5$ and $\abs{\eta^\gamma} < 1.0$, in four bins of \YoneS \pt, spanning the range 7--40\GeV.
The measurement has been performed using a data sample collected by the CMS experiment in 2012, at $\sqrt{s} = 8$\TeV, corresponding to an integrated luminosity of \usedlumi.
The cross section ratio averaged over the \YoneS \pt range is measured to be $0.85 \pm 0.07 \,\text{(stat+syst)}\pm 0.08\,\text{(BF)}$, where the first uncertainty is the combination of the experimental statistical and systematic uncertainties and the second is from the uncertainty in the ratio of the \Chib branching fractions. The ratio does not show a significant dependence on the \YoneS \pt.
This is the most precise measurement to date of the \Chitwo and \Chione relative production cross sections in hadron collisions, which complements and extends the results of Ref.~\cite{LHCB:2014chib} obtained in the kinematic region $2.0 < y(\Chib) < 4.5$,  $5.0 < \pt(\Upsilon)< 25 ~\GeV$.

\section*{Acknowledgements}
We congratulate our colleagues in the CERN accelerator departments for the excellent performance of the LHC and thank the technical and administrative staffs at CERN and at other CMS institutes for their contributions to the success of the CMS effort. In addition, we gratefully acknowledge the computing centres and personnel of the Worldwide LHC Computing Grid for delivering so effectively the computing infrastructure essential to our analyses. Finally, we acknowledge the enduring support for the construction and operation of the LHC and the CMS detector provided by the following funding agencies: BMWFW and FWF (Austria); FNRS and FWO (Belgium); CNPq, CAPES, FAPERJ, and FAPESP (Brazil); MES (Bulgaria); CERN; CAS, MoST, and NSFC (China); COLCIENCIAS (Colombia); MSES and CSF (Croatia); RPF (Cyprus); MoER, ERC IUT and ERDF (Estonia); Academy of Finland, MEC, and HIP (Finland); CEA and CNRS/IN2P3 (France); BMBF, DFG, and HGF (Germany); GSRT (Greece); OTKA and NIH (Hungary); DAE and DST (India); IPM (Iran); SFI (Ireland); INFN (Italy); NRF and WCU (Republic of Korea); LAS (Lithuania); MOE and UM (Malaysia); CINVESTAV, CONACYT, SEP, and UASLP-FAI (Mexico); MBIE (New Zealand); PAEC (Pakistan); MSHE and NSC (Poland); FCT (Portugal); JINR (Dubna); MON, RosAtom, RAS and RFBR (Russia); MESTD (Serbia); SEIDI and CPAN (Spain); Swiss Funding Agencies (Switzerland); MST (Taipei); ThEPCenter, IPST, STAR and NSTDA (Thailand); TUBITAK and TAEK (Turkey); NASU and SFFR (Ukraine); STFC (United Kingdom); DOE and NSF (USA).

Individuals have received support from the Marie-Curie programme and the European Research Council and EPLANET (European Union); the Leventis Foundation; the A. P. Sloan Foundation; the Alexander von Humboldt Foundation; the Belgian Federal Science Policy Office; the Fonds pour la Formation \`a la Recherche dans l'Industrie et dans l'Agriculture (FRIA-Belgium); the Agentschap voor Innovatie door Wetenschap en Technologie (IWT-Belgium); the Ministry of Education, Youth and Sports (MEYS) of the Czech Republic; the Council of Science and Industrial Research, India; the HOMING PLUS programme of Foundation for Polish Science, cofinanced from European Union, Regional Development Fund; the Compagnia di San Paolo (Torino); the Consorzio per la Fisica (Trieste); MIUR project 20108T4XTM (Italy); the Thalis and Aristeia programmes cofinanced by EU-ESF and the Greek NSRF; and the National Priorities Research Program by Qatar National Research Fund.

\ifthenelse{\boolean{cms@external}}{\vspace*{24pt}}{}

\bibliography{auto_generated}   
\cleardoublepage \appendix\section{The CMS Collaboration \label{app:collab}}\begin{sloppypar}\hyphenpenalty=5000\widowpenalty=500\clubpenalty=5000\textbf{Yerevan Physics Institute,  Yerevan,  Armenia}\\*[0pt]
V.~Khachatryan, A.M.~Sirunyan, A.~Tumasyan
\vskip\cmsinstskip
\textbf{Institut f\"{u}r Hochenergiephysik der OeAW,  Wien,  Austria}\\*[0pt]
W.~Adam, T.~Bergauer, M.~Dragicevic, J.~Er\"{o}, C.~Fabjan\cmsAuthorMark{1}, M.~Friedl, R.~Fr\"{u}hwirth\cmsAuthorMark{1}, V.M.~Ghete, C.~Hartl, N.~H\"{o}rmann, J.~Hrubec, M.~Jeitler\cmsAuthorMark{1}, W.~Kiesenhofer, V.~Kn\"{u}nz, M.~Krammer\cmsAuthorMark{1}, I.~Kr\"{a}tschmer, D.~Liko, I.~Mikulec, D.~Rabady\cmsAuthorMark{2}, B.~Rahbaran, H.~Rohringer, R.~Sch\"{o}fbeck, J.~Strauss, A.~Taurok, W.~Treberer-Treberspurg, W.~Waltenberger, C.-E.~Wulz\cmsAuthorMark{1}
% ...
\vskip\cmsinstskip
% ...
1:~~Also at Vienna University of Technology, Vienna, Austria\\
2:~~Also at CERN, European Organization for Nuclear Research, Geneva, Switzerland\\
% ...

\end{sloppypar}
\end{document}